\documentclass{article}

\usepackage{PRIMEarxiv}

\usepackage[utf8]{inputenc} % allow utf-8 input
\usepackage[T1]{fontenc}    % use 8-bit T1 fonts
\usepackage{hyperref}       % hyperlinks
\usepackage{url}            % simple URL typesetting
\usepackage{booktabs}       % professional-quality tables
\usepackage{amsfonts}       % blackboard math symbols
\usepackage{amsthm}         % theorem/proposition environments
\usepackage{nicefrac}       % compact symbols for 1/2, etc.
\usepackage{microtype}      % microtypography
\usepackage{textcomp}       % support Unicode minus sign
\usepackage{amssymb}        % for \geqslant and other symbols
\usepackage{lipsum}
\usepackage{fancyhdr}       % header
\usepackage{graphicx}       % graphics
\graphicspath{{media/}}     % organize your images and other figures under media/ folder
\usepackage{adjustbox}
\usepackage{xcolor}
\usepackage{float}
\usepackage{multirow}
\usepackage{amsmath}
\usepackage{natbib}

\newtheorem{proposition}{Proposition}

\title{Local Balance Calibration for Nonparametric Propensity Score Estimation
}

\author{
  Maosen Peng \textsuperscript{1,2}, 
  Yan Li \textsuperscript{3}
  Chong Wu \thanks{Corresponding author. Email: CWu18@mdanderson.org}\textsuperscript{\;\,,1},
  Liang Li\thanks{Corresponding author. Email: LLi15@mdanderson.org} \textsuperscript{\;\,,1}
  \\
  \textsuperscript{1} Department of Biostatistics, The University of Texas MD Anderson Cancer Center, Houston, Texas, USA \\
  \textsuperscript{2} Department of Biostatistics and Data Science, University of Texas School of Public Health, Houston, TX, USA\\
  \textsuperscript{3} Department of Quantitative Health Sciences, Mayo Clinic, Rochester, Minnesota, USA\\
}

\begin{document}
\maketitle

\begin{abstract}
The propensity score is widely used for causal inference in observational studies, but common parametric estimators can produce biased and inefficient effect estimates when model assumptions are violated. Nonparametric approaches reduce sensitivity to misspecification but often yield unstable weights and inadequate covariate balance. We propose Local Balance with Calibration, implemented by Neural Networks, a weighting method that combines flexible function approximation with the explicit enforcement of covariate balance and calibration. When used with inverse probability weighting, the proposed estimator produces more stable weights, improved covariate balance, and reduced bias in average treatment effect estimation compared with existing approaches. We further develop an influence-function-based variance estimator that provides accurate uncertainty quantification for the resulting weighted estimators. Numerical studies demonstrate improved efficiency and reliable variance estimation across a range of data-generating scenarios. The method is implemented using the publicly available R package \texttt{LBCNet}. 
\end{abstract}

% keywords can be removed

\section{Introduction}
\label{sec:intro}

The propensity score (PS), defined as the conditional probability of receiving treatment given covariates, is widely used in observational  studies and is also a basic building block of many causal inference methods \citep{ImbensRubinBook, RobinsBook2020}. Under appropriate assumptions, propensity score adjustment can make the observational data resemble a randomized controlled trial (RCT) by balancing confounders between treatment groups. This allows for straightforward two-sample comparisons, similar to those used in RCTs. This paper focuses on PS weighting, specifically the inverse probability of treatment weighting (IPTW), though many concepts also apply to other PS methods, such as matching.

Despite the theoretical appeal and widespread adoption of propensity scores, significant challenges persist in data analytical practice. The conventional approach employs parametric PS models, particularly logistic regression. However, model misspecification, particularly in the presence of many covariates, may lead to inadequate covariate balance, biased causal effect estimates, loss of efficiency, and instability in weights. Covariate balance checking is useful for diagnosing model misspecification \citep{austin2015moving}, but there is little theoretical guidance on how to revise a parametric model when imbalance is detected. The challenge is compounded by potential instability in IPTW weights -- even under a nearly correct PS model -- particularly when overlap is limited \citep{kang2007demystifying}. Such instability can bias balance measures, further complicating model diagnosis. Estimation under covariate balance constraints can improve finite-sample performance and reduce weight instability \citep{imai2014covariate}. However, enforcing balance with a misspecified model form may mask underlying issues, rendering balance checks ineffective \citep{li2021propensity}. 

Replacing the parametric model with nonparametric machine learning methods, while reducing concerns about model misspecification, does not inherently guarantee covariate balance or robustness \citep{Setoguchi2008, lee2010improving, cannas2019comparison}. This stems from a fundamental misalignment of objectives: most machine learning methods are designed to maximize predictive accuracy, whereas unbiased estimation of treatment effects requires minimizing covariate imbalance \citep{shang2025robust}. To address this limitation, \citet{zhao2019covariate} designed a loss function, i.e., the covariate balance scoring rule (CBSR), for nonparametric PS estimation. The CBSR loss quantifies the mean imbalance in covariates. It performed better than off-the-shelf nonparametric regression models. However, the weights remain variable, contributing to a loss in efficiency \citep{li2021propensity}; closer examination reveals a notable covariate imbalance across regions of the propensity scores, suggesting that the loss function remains inadequate in constraining the weights to produce the desired balancing properties of the propensity score \citep{PSLB2023}. \citet{shang2025robust} further proposed a calibrated loss function that augments the Bernoulli log-likelihood with a covariate imbalance penalty, thereby balancing predictive accuracy and global covariate balance. Nevertheless, since the balancing objective is imposed only through penalization, the resulting estimator does not always achieve good balancing and estimation performance, as demonstrated in our numerical studies.

To address these challenges, we introduce a novel loss function for PS estimation. The loss is grounded in a novel utilization of the propensity score based on two conditions that are jointly necessary and sufficient for a function of covariates to be a valid propensity score: (1) local balance, which ensures the conditional independence of covariates and treatment given the score and is the balancing property of the propensity score as established by \citet{rosenbaum1983central}, and (2) local calibration, which ensures that the score is a true conditional probability of treatment. Together, these conditions offer an alternative definition of the propensity score. The derived weights are not merely an instrument for achieving covariate balance but also have a meaningful interpretation as the IPTW, and the estimated treatment effect retains its interpretation as the average treatment effect. By directly quantifying covariate imbalance, the loss function achieves superior balance compared to standard machine learning approaches. While \citet{PSLB2023} also used local balance to produce promising results, their two-step procedure is approximate and lacks a strong theoretical foundation; without the local calibration constraint, it produces balancing scores instead of the propensity scores.

Another line of research focuses on flexible weighting methods, often developed outside the PS framework, that directly optimize discrepancies in covariate distributions between treatment groups \citep{WongChan2017, hazlett2018kernel, Chattopadhyay2020, benmichael2021balancing, FanImai2021, Covariate.distribution.balance.2022, kong2023covariate, KimZubizarreta2023, Keele2025}. Many of these approaches build upon earlier developments that rely on a finite set of prespecified balance constraints \citep{ hainmueller2012entropy, imai2014covariate, zubizarreta2015stable}, which share conceptual connections with parametric PS estimation \citep{WangZubiz2019}. We refer to these as global covariate balance constraints (see Section \ref{sec:method}), as they minimize overall covariate differences between treatment groups. In contrast, our method is motivated by local balance, which focuses on differences within PS strata. Local balance implies global balance, but the converse does not hold. We include some recent global balance methods in our comparative numerical studies.

In addition to point estimation, valid uncertainty quantification for inverse probability weighting estimators with nonparametric PS estimation remains challenging. Unlike doubly robust estimators, for which efficient influence-function-based variance estimators are well established under semiparametric theory \citep{vandervaart1998, newey1994asymptotic}, there are limited general-purpose methods for variance estimation when inverse probability weighting is combined with flexible, high-dimensional PS models. Although influence functions provide a principled characterization of asymptotic variability, their practical use depends critically on the stability and regularity of the nuisance estimators. Recent work has extended influence-function-based (IF-based) ideas to deep neural networks by studying the effects of infinitesimal parameter perturbations on predictive functionals \citep{koh2017}. Building on these developments, we develop an IF-based variance estimator tailored to inverse probability weighting with nonparametric PS estimation without the use of an outcome model, bridging classical semiparametric theory and modern neural network–based estimation.

The proposed estimation algorithm, named LBC-Net (Local Balance with Calibration implemented by Neural Networks), can be conveniently implemented using the widely available deep learning platform, with minimal manual tuning. The R package is available at \url{https://maosenpeng1.github.io/LBCNet/}. The rest of the paper is organized into four sections: methodology presentation (Section \ref{sec:method}), simulation (Section \ref{sec:simulation}), illustrative data application (Section \ref{sec:application}), and discussion (Section \ref{sec:discussion}). Comprehensive numerical studies are essential for establishing the validity of the proposed methodology. Due to the page limit, most of them are placed in the online supplementary materials, accessible through the journal's website, and only selected results are  included in Sections \ref{sec:simulation} and \ref{sec:application}.

\section{The Proposed Methodology}
\label{sec:method}

We consider a sample of $N$ subjects indexed by $i$, comprising $N_0$ untreated ($T_i=0$) and $N_1$ treated ($T_i=1$) subjects. For subject $i$, we observe a vector of $M$ covariates $\mathbf{Z}_i = (Z_{i1}, \ldots, Z_{iM})^\top$ and the outcome variable $Y_i$. The propensity score is $p_i = p(\mathbf{Z}_i) = P(T_i=1 \mid \mathbf{Z}_i)$. We adopt the standard assumptions for PS analysis. These include the stable unit treatment value assumption, which ensures no unmodeled spillovers; the strong ignorability assumption, implying no unmeasured confounders; and the overlap assumption, ensuring that the propensity score is bounded away from 0 or 1. Under these conditions, the propensity score exhibits the balancing property $ T_i \perp \mathbf{Z}_i \mid p(\mathbf{Z}_i) $. The analytical goal is to estimate the frequency-weighted average of individual treatment effects over the sampling population, given by: 
\begin{equation}
    \Delta = \frac{E(g(p_i)\Delta_i)}{E(g(p_i))} .\notag 
\end{equation}
where $\Delta_i = E[ Y_i(1) - Y_i(0) \mid \boldsymbol Z_i]$ is the individual conditional treatment effect for subject $i$. Here, $g(.)$ is a frequency weight function of the propensity score proposed by \citet{Mao2018}, which determines the target population for treatment effect estimation. For example, setting $g(p_i)=1$ yields the Average Treatment Effect (ATE), while choosing $g(p_i)=p_i$ results in the Average Treatment Effect on the Treated (ATT). In this paper, all numerical studies focus on the ATE.

We use general IPTW weights to estimate the treatment effect, where the weight for subject $i$ is defined as:
\begin{equation}
    W_i = \dfrac{g(p_i)}{ T_i p_i + (1-T_i)(1-p_i) } .\notag 
\end{equation}
\label{eq:ipw_hajek}
The IPTW estimator of the ATE is motivated by the equation $\Delta = E( T_i W_i Y_i ) - E[ (1-T_i) W_i Y_i ]$. A consistent H\'ajek estimator of $\Delta$ can be formed as :
\begin{equation}
    \hat{\Delta} = \dfrac{\sum_{i=1}^N T_i W_i Y_i}{\sum_{i=1}^N T_i W_i} - \dfrac{\sum_{i=1}^N (1-T_i) W_i Y_i}{\sum_{i=1}^N (1-T_i) W_i}  .
\end{equation}

\subsection{Sufficient and necessary conditions for the propensity scores}
\label{subsec:nonpar.PS}

To establish a rigorous foundation for nonparametric PS estimation, we consider a mapping $S = S(\boldsymbol Z)$ from a set of covariates $\boldsymbol Z$ to a probability $S$, where $S \in (0,1)$. To simplify the notation, we omit the subscript $i$ here when referring to random variables of a generic subject. Our objective is to delineate the specific conditions under which $S$ qualifies as a propensity score. A natural choice is the standard definition $S = E( T | \boldsymbol Z )$. This mapping can be learned through various binary models of regressing $T$ on $\boldsymbol Z$, such as logistic regression, random forests, or neural networks. However, parametric models risk misspecification, while nonparametric models do not guarantee covariate balance, both of which could result in biased treatment effect estimates \citep{kang2007demystifying, li2021propensity}. We address these limitations by establishing the following characterization:

%\begin{theorem}
\begin{proposition}
\label{thm:two.conditions} The sufficient and necessary conditions for $S = p$ are: (1) the local balance condition, $T \perp \boldsymbol Z | S $, and (2) the local calibration condition, $S = E( T | S )$. 
\end{proposition}
%\end{theorem}
\begin{proof} 
For the necessary part: When $S=p$, conditions (1) and (2) hold by the properties of the propensity score. For the sufficiency part: according to Theorem 2 of \citet{rosenbaum1983central}, condition (1) defines $S$ as a balancing score, and the propensity score $p$ must be a function of $S$. Let this function be $g(.)$ such that $p = g(S)$. From condition (2) and the law of total expectation, we have $S = E[ E( T | \boldsymbol Z, S) | S ] = E[ E( T | \boldsymbol Z ) | S ] = E( p | S )$. Since $S = E( g(S) | S ) = g(S)$, we have $p = S$.  
\end{proof}
This result provides two key insights. First, the local balance condition ensures that $S$ functions as a balancing score; however, this property alone does not guarantee that it captures the true treatment assignment mechanism. Second, the local calibration condition maps a balancing score to a proper propensity score. Together, these conditions offer a novel perspective on PS estimation: rather than modeling the treatment assignment mechanism directly, valid propensity scores can be identified by searching for mappings that simultaneously satisfy these two properties. The resulting IPTW estimator targets the average treatment effect.

Proposition \ref{thm:two.conditions} can be viewed as an alternative definition of the propensity score. The standard definition, i.e., the conditional distribution of $T | \boldsymbol Z$, does not directly address covariate balance. In contrast, the propensity scores satisfying the alternative definition are expected to produce good local balance; local balance at all PS strata implies good global balance—the prevalent measure for checking covariate balance in the PS literature (\textbf{Remark 2.3}). In numerical studies, we have observed that the loss based on this alternative definition tends to achieve superior covariate balance compared to the loss function based on the standard definition, such as the cross-entropy (equivalent to the log-likelihood of a Bernoulli distribution). While we focus our comparisons on cross-entropy loss, similar concerns apply to classification error loss, which also prioritizes predictive accuracy over covariate balance. Therefore, we suggest that if a machine learning method for binary outcomes is to be used for PS estimation, the conventional loss that quantifies prediction error is not desirable; the tailored loss function in the next section is a better choice. 

\subsection{Constructing the loss function}
\label{subsec:target.function}

To operationalize the conditions established in Proposition \ref{thm:two.conditions}, we develop a tractable optimization framework for finding the nonparametric mapping $\boldsymbol Z: \rightarrow S$ that meets the sufficient and necessary conditions for $S = p$. For this purpose, we construct empirically verifiable equations from the local balance condition $E( T W \boldsymbol Z | S ) - E[ (1-T) W \boldsymbol Z | S ] = \boldsymbol 0 $ and the local calibration condition $E( T - S | S) = 0$. To verify these conditions empirically, we evaluate them on a dense grid of points spanning $(0,1)$, denoted by $c_k = k/(K+1)$, where $k=1,2,\ldots,K$. Each $c_k$ defines a local neighborhood. The observations within a neighborhood are weighted by kernel weights. Specifically, for $c_k$, we define Gaussian kernel weights $\omega(c_k, x) = h_k^{-1} K\left[ (c_k - x)/h_k \right]$, where  $K(x) = (2\pi)^{-1/2} \exp\left( -x^2/2 \right)$ and $h_k$ represent location-specific bandwidths. Other kernel functions can also be used. This construction leads to a local inverse probability of treatment weight for subject $i$, defined as $W_k(p_i)=\omega(c_k, p_i)W_i$. For the ATE, the corresponding local weight takes the form:  
\begin{equation}
    W_k(p_i) = \dfrac{ \omega(c_k, p_i) }{ T_i p_i + (1-T_i)(1 - p_i) }  ~~,~~ i = 1, 2, ..., N. \notag 
\end{equation} 
where $p_i = p_\theta(\mathbf Z_i)$ and $\boldsymbol\theta$ denote the vector of trainable parameters indexing the PS model.

Compared with the IPTW weight, the difference lies in the numerator. The balancing weights \citep{LiFan2017JASA}, such as the matching weight \citep{li2013weighting}, have similarly formulated weights; however, the key difference is that these methods first estimate the propensity scores using the standard definition and then substitute the estimates into the weights formula. In this paper, however, the propensity scores $p_i$ in both the numerator and denominator are unknown and are estimated using a deep neural network, as described in Section \ref{subsec:PSLB-DL}.  

Using these weights, we construct the sample version of the local balance equation:
\begin{equation}
\begin{aligned}
    \boldsymbol D_{1k} =  ~ \sum^{N}_{i=1}T_i W_k(p_i) \boldsymbol Z_i - \sum^{N}_{i=1}(1 - T_i) W_k(p_i) \boldsymbol Z_i \\  
             =  ~ \sum^{N}_{i=1} \omega(c_k, p_i) \boldsymbol V_i , \notag 
\end{aligned}
\end{equation} for $k=1,2,..., K$, where $\boldsymbol V_i = \dfrac{ (2T_i-1) \boldsymbol Z_i }{ T_i p_i + (1-T_i)(1-p_i) }$. Under a correctly specified PS model, $E( \boldsymbol D_{1k} ) = \sum_{i=1}^N E\left[  \omega(c_k, p_i) \boldsymbol V_i \right] = \boldsymbol 0$ follows from the fact that $E( \boldsymbol V_i | p_i ) = \boldsymbol 0$ and $E\left[ \omega(c_k, p_i) \boldsymbol V_i \right] = E\left[ \omega(c_k, p_i) E( \boldsymbol V_i | p_i ) \right]$. To account for varying scales among covariates, we define a standardization matrix: 
\begin{equation}
\label{eq:SigmaDef}
\begin{aligned}
\boldsymbol \Sigma_{k}
&= E\!\left( \boldsymbol D_{1k} \boldsymbol D_{1k}^\top \right) \\
&= E\left\{ \sum_{i=1}^N \omega(c_k, p_i)^2 \boldsymbol V_i \boldsymbol V_i^\top \right\} \\
&= E\left\{ \sum_{i=1}^N
\frac{ \omega(c_k, p_i)^2 \boldsymbol Z_i \boldsymbol Z_i^\top }
{ T_i p_i^2 + (1-T_i)(1-p_i)^2 }
\right\} \\
&= E\left\{ \sum_{i=1}^N
\frac{ \omega(c_k, p_i)^2 \boldsymbol Z_i \boldsymbol Z_i^\top }
{ p_i(1-p_i) }
\right\}.
\end{aligned}
\end{equation}
In practice, we approximate $\boldsymbol \Sigma_k$ using the sample-based estimator
\begin{equation}
\widehat{\boldsymbol \Sigma}_k
=
\frac{1}{c_k(1-c_k)}
\sum_{i=1}^N
\omega(c_k,p_i)^2 \boldsymbol Z_i \boldsymbol Z_i^\top, \notag
\end{equation}
where the role of $\Sigma_k$ and its relationship to the moment-based representation is discussed in Section~\ref{subsec:gmm.function}.

Given these components, we obtain the first component of the loss:  
\begin{equation}
    Q_1(\boldsymbol \theta) = \dfrac{1}{K}\sum_{k=1}^K \boldsymbol D_{1k}^\top \boldsymbol \Sigma_k^{-1} \boldsymbol D_{1k} ~. \notag 
\end{equation}
By default, we include an intercept term of $1$ in $\boldsymbol Z$ to ensure that $\sum_{i=1}^N T_i W_k(p_i) \approx \sum_{i=1}^N (1-T_i) W_k(p_i) $ encodes a relationship that is expected to hold in each local neighborhood under the correct PS model. In summary, $Q_1(\boldsymbol \theta)$ aggregates the contributions across all $K$ local neighborhoods, providing an overall measure of local balance. 

For the local calibration, we construct a standardized measure at each grid point:
\begin{equation}
    D_{2k} = \sum_{i=1}^N \omega(c_k, p_i) \left( \dfrac{ T_i - p_i }{ \sqrt{ c_k (1 - c_k) } } \right) ~ . \notag   
\end{equation}
The expectation of $D_{2k}$ is zero if the PS model is correctly specified. The denominator $\sqrt{ c_k (1 - c_k) }$ draws a connection with the $\chi^2$ statistic and ensures approximately equal contributions across the PS range, avoiding down-weighting local neighborhoods near $0$ or $1$. This leads to the second  component of the loss, the $L_2$ norm of $D_{2k}$ summed over K:  
\begin{equation}
\begin{aligned}
    Q_2(\boldsymbol \theta) = & ~ \dfrac{1}{K} \sum_{k=1}^K \left( \dfrac{ \sum_{i=1}^N \omega(c_k, p_i)( T_i - p_i )^2 }{ c_k(1-c_k) \sum_{i=1}^N \omega(c_k, p_i) } \right) ~~,~~\notag 
\end{aligned}
\end{equation}
which is analogous to the Hosmer-Lemeshow goodness-of-fit statistic. 

The loss function synthesizes both local balance and local calibration components: $Q(\boldsymbol \theta) = Q_1(\boldsymbol \theta) + \lambda Q_2(\boldsymbol \theta)$, where the weight $\lambda \geqslant 0$ controls their relative contributions. This synthesis is referred to as a multi-head loss function in machine learning. The following proposition characterizes the asymptotic behavior of the two components, which helps justify $\lambda$. 

% \begin{theorem}
\begin{proposition}
\label{thm:Q.limit}
Suppose that the following regularity conditions hold: (1) the kernel function $K(\cdot)$ has bounded support, (2) $K \rightarrow \infty$ and $K/N \rightarrow 0$, (3) $ max\{ 2h_k \} \leqslant 1/(K+1) $, and (4) the density function of the propensity score is bounded away from $0$ on its support. With the true propensity scores, we have $\underset{N \rightarrow \infty}{\mathrm{lim}} Q_1(\boldsymbol \theta) = L$, where $L$ is the dimension of $\boldsymbol Z$, and $\underset{N \rightarrow \infty}{\mathrm{lim}} Q_2(\boldsymbol \theta) = 1$. 
\end{proposition}
%\end{theorem}
\begin{proof} 
$\boldsymbol D_{1k} = N \times \dfrac{\sum_{i=1}^N \omega(c_k, p_i) }{N} \times \dfrac{ \sum_{i=1}^N \omega(c_k, p_i) \boldsymbol V_j }{ \sum_{i=1}^N \omega(c_k, p_i) } $. The middle term converges to $f_p(c_k)$, the density function of the propensity score evaluated at $c_k$. The third term is a Nadaraya-Watson estimator of $\boldsymbol V$. Therefore, $\boldsymbol D_{1k}$ is asymptotically normal. Since $E( \boldsymbol D_{1k} ) = \boldsymbol 0$ and $\boldsymbol \Sigma_{k} = E( \boldsymbol D_{1k} \boldsymbol D_{1k}^\top )$, $\boldsymbol D_{1k}^\top \boldsymbol \Sigma_k^{-1} \boldsymbol D_{1k}$ has an asymptotically central chi-square distribution with $L$ degrees of freedom. The regularity conditions imply that $Q_1(\boldsymbol \theta)$ is the average of $K$ independent terms. As $K \rightarrow \infty$, the limit of $Q_1(\boldsymbol \theta)$ is $L$. Since $ \dfrac{ \sum_{i=1}^N \omega(c_k, p_i)( T_i - p_i )^2 }{ \sum_{i=1}^N \omega(c_k, p_i) } $ is the Nadaraya-Watson estimator of $\mathrm{var}( T_i | p_i = c_k ) = c_k(1-c_k)$, $Q_2(\boldsymbol \theta)$ is the average of $K$ independent terms, each with a mean of $1$. As $K \rightarrow \infty$, the limit of $Q_2(\boldsymbol \theta)$ is $1$. 
\end{proof}
Proposition~\ref{thm:Q.limit} suggests that when the estimated propensity scores approach the truth, the components in $Q(\boldsymbol \theta)$ contribute in a ratio of $L/\lambda$. The regularity conditions in Proposition~\ref{thm:Q.limit} represent the ideal situation and may not all hold in data analysis practice. Since our goal is to develop an automated PS analysis tool with minimal human operation,  we recommend $\lambda = 1$ as the default. In our numerical studies, as well as in some other literature using the multi-head loss function \citep{2019dragonnet}, the results are not sensitive to moderate variations in $\lambda$. 

\textbf{Remark 2.1}. Although the loss derivation in this section focuses on the estimation of the ATE, the extension to the ATT estimand is straightforward and is implemented in our \textsf{R} package.

\textbf{Remark 2.2}. The covariate vector $\boldsymbol Z$ may include arbitrary pre-specified transformations of the original covariates, such as nonlinear terms or interactions, without altering the proposed methodology or computational algorithm. This flexibility allows the balance conditions to target higher-order features of the covariate distribution when desired.

\textbf{Remark 2.3}. The local balance condition implies global balance. Specifically, taking the expectation with respect to $p$ on both sides of $E( T W \boldsymbol Z | p ) - E[ (1-T) W \boldsymbol Z | p ] = \boldsymbol 0 $ yields $E( T W \boldsymbol Z ) - E[ (1-T) W \boldsymbol Z ] = \boldsymbol 0$, the global balance. However, the converse does not hold; global balance alone cannot guarantee local balance. This distinction is important, as methods that force global balance without achieving local balance may mask underlying model misspecification and cause the bias of the ATE to go unnoticed \citep{li2021propensity}. 

\subsection{Moment-based representation and estimation}
\label{subsec:gmm.function}

Although LBC-Net is introduced through a loss function, the resulting estimator admits an equivalent representation in terms of estimating equations, which can be expressed within a generalized method of moments (GMM) framework. The loss formulation presented in Section~\ref{subsec:target.function} provides a direct and intuitive description of the estimation procedure and has been validated in
earlier versions of this work \citep{anonlbc}. In the present paper, we exploit the simplified but equivalent moment-based representation, which is particularly convenient for variance estimation in Section \ref{sec:variance}. All numerical results reported herein are computed using this formulation.

Let $\boldsymbol X_i=(T_i, \boldsymbol Z_i)$. The parameter vector $\boldsymbol\theta$ indexing the PS model $p_\theta(\mathbf X)$ is characterized as the solution to the empirical moment conditions in the last subsection that enforce local balance and local calibration over a grid of kernel locations $\{(c_k,h_k)\}_{k=1}^K$. We consider the default setting with $\lambda = 1$ and adopt the identity weighting matrix $\boldsymbol \Sigma_k= I$. All covariates are standardized prior to optimization, and the primary objective is covariate balance rather than asymptotic efficiency. Under these conditions, identity weighting induces a stable and interpretable geometry for the moment conditions. This yields a GMM formulation with identity weighting.

For each kernel node $c_k$, define the local moment vector
\begin{equation}
m_k(\boldsymbol X_i,\boldsymbol\theta)
=
\omega(c_k,p_i)
\begin{pmatrix}
V_i^\top\\
\dfrac{T_i-p_i}{c_k(1-c_k)}
\end{pmatrix}, \notag 
\end{equation}
where $p_i = p_\theta(\boldsymbol X_i)$. Then, stacking the $K$ local moment vectors yields 
\begin{equation}
m(\boldsymbol X_i,\boldsymbol\theta)
=
\left(
m_1(\boldsymbol X_i,\boldsymbol\theta)^\top,\,
\ldots,\,
m_K(\boldsymbol X_i,\boldsymbol\theta)^\top
\right)^\top, \notag
\end{equation}
where $m(\boldsymbol X_i,\boldsymbol\theta) \in \mathbb{R}^{K(M+1)}$. Let $\boldsymbol\theta_0$ denote the population parameter satisfying
$\mathbb{E}_{P_0}\{ m(\boldsymbol X,\boldsymbol\theta_0) \}=0$, where $P_0$ denotes the true data-generating distribution of $\boldsymbol X$. Define the empirical moment vector $\hat m_N(\boldsymbol\theta) = \frac{1}{N}\sum_{i=1}^N m(\boldsymbol X_i,\boldsymbol\theta)$; then the corresponding quadratic GMM loss function of the LBC-Net estimator is
\begin{equation}
Q^\ast(\boldsymbol\theta)
=
\frac{1}{K}\,
\hat m_N(\boldsymbol\theta)^\top
\hat m_N(\boldsymbol\theta).
\label{eq:newloss}
\end{equation}

The next section describes the computational algorithm used to minimize $Q^\ast(\boldsymbol\theta)$, including the optimization scheme and key implementation details required for reproducibility. 

\subsection{Optimizing the loss function with LBC-Net}
\label{subsec:PSLB-DL}

Since the algorithm optimizes local balance and local calibration, it is necessary for users to define locality as input to the algorithm. That includes $\{ (c_k, h_k) ; k=1,2,..., K\}$, the mid-point and bandwidth of each local neighborhood. With equally spaced mid-points that span the whole range from 0 to 1, they can be determined by $K$, the number of local neighborhoods. We consider a fine grid with $K=99$ and $\{ c_k \} = 0.01, 0.02, ..., 0.99$ as the default option, and the robustness of results to $K$ is explored in simulations (Section \ref{sec:simulation}). 

We recommend using adaptive rather than fixed bandwidths, as the latter can lead to small sample sizes in regions with sparse propensity scores, resulting in spurious covariate imbalances. Each bandwidth $h_k$ is determined from a user-defined span $\rho \in (0,1)$, which specifies the proportion of data included in each local neighborhood. For a given $\rho$, the bandwidth $h_k$ is chosen as the smallest value such that the neighborhood centered at $c_k$ contains a fraction $\rho$ of the sample.

Because the true PS distributions are unknown, we approximate it using propensity scores estimated from a preliminary logistic regression with main effects of covariates. This step is used solely to determine the geometry of the local neighborhoods: only the shape of the propensity score distribution is required to map $\rho$ to the collection $\{h_k\}$, not the exact score values. The resulting bandwidths are then treated as inputs to the objective function.

Our practical guidance for choosing $\rho$ is as follows. Moderate spans, typically in the range $\rho \in [0.15, 0.25]$, provide stable performance across a wide range of settings, with $\rho = 0.15$ serving as a default that consistently achieves good local balance. Smaller values improve locality but can lead to unstable moment estimates in finite samples, whereas larger values sacrifice local adaptivity in favor of global balance. For small samples or limited overlap, we recommend increasing $\rho$ (e.g., to $0.20$ or $0.25$) to ensure adequate local sample sizes. The Supplementary Material reports sensitivity analyses across sample sizes and overlap regimes, demonstrating that covariate balance and treatment effect estimates are stable within this recommended range of $\rho$ (Tables S10, S13, and Figure S28).

Our goal is to reduce the amount of user intervention required for implementing the proposed procedure. At the methodological level, the algorithm requires two primary user-specified inputs, namely, the number of local balance points $K$ and the neighborhood span $\rho$, both of which admit default values that perform well across a wide range of numerical settings. Throughout our experiments, we use the same $\{ c_k \}$ and $\{ h_k \}$ in both the loss function and the balance diagnostics (Section~\ref{subsec:GSD.LSD}). We further evaluate local balance at midpoints between adjacent $c_k$'s and observe comparable balance levels, indicating that local balance varies smoothly with respect to the propensity score and that enforcing balance at a moderate number of representative points also controls imbalance in between. As a result, the method is generally insensitive to the precise choice of $K$, provided that adjacent local neighborhoods exhibit sufficient overlap.

Minimizing the loss function $Q^\ast(\boldsymbol\theta)$ is computationally nontrivial because the objective depends on localized weights $W_k(p_i)$ that vary with the current PS estimates. We implement the minimization using a flexible feed-forward neural network, which provides a stable and scalable parameterization for the PS surface under the proposed balance-driven objective. To improve initialization and training stability, we use a variational auto-encoder for pre-training. Training is terminated using an early stopping criterion based on covariate balance diagnostics. Detailed specifications of the neural network architecture, optimization settings, and stopping criteria are provided in the Supplementary Material (Section S3).

As with any neural network implementation, additional training hyperparameters (e.g., learning rate, network width and depth, and regularization) must be specified. In the present setting, these choices are less critical than in prediction-oriented
applications, as the network is trained to minimize covariate imbalance rather than to optimize predictive performance. Sensitivity analyses in the Supplementary Material indicate that within a reasonable range, these hyperparameters have only a marginal impact on covariate balance and treatment effect estimation, with default settings performing well in most scenarios (Tables S8–10, S13–15; Figures S13-14 and S17-18); in particular, a three-layer architecture provides consistently robust performance.

\subsection{Balance measures}
\label{subsec:GSD.LSD}

Covariate balance assessment is central to evaluating the adequacy of PS models. We examine both global and local balance measures, as deficiencies in either dimension may indicate model misspecification and reduce confidence in the resulting treatment effect estimates. For global balance, we employ the standardized mean difference (GSD), expressed as a percentage. For a generic scalar covariate $Z_{1}$ or its transformation, the GSD is defined as
\begin{equation}
\mathrm{GSD}
=
\frac{|\bar{z}_1 - \bar{z}_0|}
{\sqrt{(m_1 v_1 + m_0 v_0)/(m_1 + m_0)}} \times 100\%, \notag
\end{equation}
where $\bar{z}_1 = \sum_{i=1}^N T_i W_i Z_{1i} / \sum_{i=1}^N T_i W_i$ and
$\bar{z}_0 = \sum_{i=1}^N (1-T_i) W_i Z_{1i} / \sum_{i=1}^N (1-T_i) W_i$
are the inverse probability weighted means for the treated and untreated groups. The corresponding weighted variances are
$v_1 = \sum_{i=1}^N T_i W_i (Z_{1i} - \bar{z}_1)^2 / (\sum_{i=1}^N T_i W_i - 1)$
and
$v_0 = \sum_{i=1}^N (1-T_i) W_i (Z_{1i} - \bar{z}_0)^2 / (\sum_{i=1}^N (1-T_i) W_i - 1)$.
The quantities $m_1$ and $m_0$ denote the effective sample sizes for the weighted treated and control groups \citep{mccaffrey2004propensity}, defined as
$m_1 = (\sum_{i=1}^N T_i W_i)^2 / \sum_{i=1}^N T_i W_i^2$
and
$m_0 = (\sum_{i=1}^N (1-T_i) W_i)^2 / \sum_{i=1}^N (1-T_i) W_i^2$.
In practice, $m_1 = m_0$ is often used as a simplifying approximation \citep{li2013weighting}.

To assess local covariate balance, we introduce a local analog of the standardized mean difference, referred to as the local standardized difference (LSD). For pre-specified grid points $c_k$, $k=1,\ldots,K$, local balance is evaluated using the kernel-based inverse probability weights $W_k(p_i)$ defined in Section~\ref{subsec:target.function}. The LSD at location $c_k$, denoted by $\mathrm{LSD}(c_k)$, is computed using the same functional form as the GSD, with the global weights replaced by the corresponding local weights $W_k(p_i)$. While the GSD can be computed for a broad class of weighting estimators, including methods that do not rely on propensity scores, the LSD is specific to PS–based approaches, as it leverages local neighborhoods defined along the estimated PS scale. In our data application, we also used empirical cumulative distribution functions to compare the differences in the covariate distribution rather than the mean difference of covariates (or transformations). 

\section{Variance Estimation}
\label{sec:variance}

In this section, we develop an influence-function–based variance estimator for IPTW estimators that use propensity scores learned via LBC-Net. Sampling variability arises from both empirical fluctuations in outcomes and treatment assignments with a fixed propensity score, and from the estimation of the propensity score itself. The latter component is obtained by linearizing the Generalized Method of Moments (GMM) estimator (\citealp{newey1994asymptotic}) underlying LBC-Net and propagating this uncertainty to the treatment effect estimator via a first-order chain-rule expansion.

We make the dependence on the neural network parameters $\boldsymbol\theta$ explicit in the notation. Let $\Delta(\boldsymbol\theta)$ denote the population IPTW H\'ajek estimand evaluated at the propensity score $p_{\boldsymbol\theta}(\boldsymbol X)$. Let $\hat{\Delta} = \Delta_N(\hat{\boldsymbol\theta})$ denote the corresponding sample IPTW estimator obtained by replacing $\boldsymbol\theta$ with its LBC-Net estimate $\hat{\boldsymbol\theta}$, which is estimated from a sample of $N$ units. To characterize the sampling variability of $\hat{\Delta}$, we introduce the auxiliary quantity $\Delta_N(\boldsymbol\theta_0)$, the sample IPTW estimator computed using the true propensity score parameter $\boldsymbol\theta_0$. Although not observable in practice, it isolates sampling variability when the propensity score is treated as fixed. Adding and subtracting this term yields the decomposition
\begin{equation}
\hat{\Delta}-\Delta(\boldsymbol\theta_0)
=
\Bigl\{\Delta_N(\boldsymbol\theta_0)-\Delta(\boldsymbol\theta_0)\Bigr\}
+
\Bigl\{\Delta_N(\hat{\boldsymbol\theta})-\Delta_N(\boldsymbol\theta_0)\Bigr\}.
\label{eq:decomp_basic}
\end{equation}

This decomposition naturally separates the asymptotic variability into two components. The first term corresponds to the empirical fluctuation of the IPTW H\'ajek estimator when the propensity score is treated as fixed, while the second term captures the additional variability induced by estimating the propensity score. As for the first component, within the fixed-propensity-score semiparametric framework of \citet{hahn1998propensity}, the influence function of the IPTW estimator corresponds to the special case of the general ATE influence function in which the outcome regression functions are constant. A first-order normalization in the Hájek form then gives the influence function that admits an asymptotically linear expansion at a generic observation $O_i=(Y_i, T_i, \boldsymbol X_i)$:
\begin{equation}
\mathrm{IF}_{\mathrm{\Delta\mid \boldsymbol\theta_0}}(\boldsymbol O_i)
=
\frac{T_i Y_i/p_{\theta_0}(\boldsymbol X_i)-\mu_1\,T_i/p_{\theta_0}(\boldsymbol X_i)}
{\mathbb{E}\{T/p_{\theta_0}(\boldsymbol X)\}}
-
\frac{(1-T_i) Y_i/(1-p_{\theta_0}(\boldsymbol X_i))-\mu_0\,(1-T_i)/(1-p_{\theta_0}(\boldsymbol X_i))}
{\mathbb{E}\{(1-T)/(1-p_{\theta_0}(\boldsymbol X))\}}, \notag
\end{equation}
where $\mu_1=\mathbb{E}\{Y(1)\}$ and $\mu_0=\mathbb{E}\{Y(0)\}$ denote the marginal mean potential outcomes under treatment and control, respectively. 

As for the second term in \eqref{eq:decomp_basic}, the parameter
$\hat{\boldsymbol\theta}$ is defined as the minimizer of the quadratic GMM criterion introduced in Section~\ref{subsec:gmm.function}. Let $G_0=\frac{\partial}{\partial \boldsymbol\theta^{\top}}\mathbb{E}_{P_0}\!\left\{ m(X,\boldsymbol\theta) \right\}|_{\boldsymbol\theta=\boldsymbol\theta_0}$ denote the Jacobian matrix of the population moment condition in Section \ref{subsec:gmm.function}. Under standard regularity
conditions for GMM estimation \citep{newey1994asymptotic}, a first-order
linearization of the first-order condition $\nabla_{\boldsymbol\theta}Q^*(\hat{\boldsymbol\theta})=0$ implies the asymptotically linear representation \citep{ichimura2022influence}
\begin{equation}
\sqrt{N}(\hat{\boldsymbol\theta}-\boldsymbol\theta_0)
=
\frac{1}{\sqrt{N}}\sum_{i=1}^N IF_{\boldsymbol\theta = \boldsymbol\theta_0}(\boldsymbol X_i)
+ o_p(1),
\qquad
IF_{\boldsymbol\theta = \boldsymbol\theta_0}(\boldsymbol X_i)
=
-\bigl(G_0^{\!\top} G_0\bigr)^{-1}
G_0^{\!\top}
m(\boldsymbol X_i,\boldsymbol\theta_0). \notag
\end{equation}
Thereafter, assuming that $\Delta_N(\boldsymbol\theta)$ is differentiable with
respect to $\boldsymbol\theta$ in a neighborhood of $\boldsymbol\theta_0$, a
first-order Taylor expansion yields
\begin{equation}
\Delta_N(\hat{\boldsymbol\theta})-\Delta_N(\boldsymbol\theta_0)
=
\left.
\frac{\partial \Delta_N(\boldsymbol\theta)}{\partial \boldsymbol\theta^{\top}}
\right|_{\boldsymbol\theta=\boldsymbol\theta_0}
(\hat{\boldsymbol\theta}-\boldsymbol\theta_0)
+o_p(N^{-1/2}). \notag
\end{equation}
Finally, combining the above results, the estimator $\hat{\Delta}$ admits an
asymptotically linear representation, whose influence function is given by
\begin{equation}
\mathrm{IF}_{\Delta}(\boldsymbol O_i)
=
\mathrm{IF}_{\mathrm{\Delta\mid \boldsymbol\theta_0}}(\boldsymbol O_i)
+
\left.
\frac{\partial \Delta(\boldsymbol\theta)}{\partial \boldsymbol\theta^{\top}}
\right|_{\boldsymbol\theta=\boldsymbol\theta_0}
\mathrm{IF}_{\boldsymbol\theta = \boldsymbol\theta_0}(\boldsymbol X_i).
\label{eq:if_delta_decomp}
\end{equation}
Since $\Delta_N(\boldsymbol\theta)$ is a smooth sample analog of
$\Delta(\boldsymbol\theta)$, the population derivative above can be consistently
estimated by the corresponding sample derivative. Under standard regularity conditions ensuring $\sqrt{N}$-consistency and asymptotic normality \citep{newey1994asymptotic,vandervaart1998}, the estimator
$\hat{\Delta}$ admits an asymptotically linear representation. Specifically, $\sqrt{N}\bigl\{\hat{\Delta}-\Delta(\boldsymbol\theta_0)\bigr\}=\frac{1}{\sqrt{N}}\sum_{i=1}^N\mathrm{IF}_{\Delta}(\boldsymbol O_i)+ o_p(1)$ and by the central limit theorem $\sqrt{N}\bigl(\hat{\Delta}-\Delta(\boldsymbol\theta_0)\bigr)
\xrightarrow{d}\mathcal N\!\left(0,\,\mathbb{E}\!\left[\mathrm{IF}_{\Delta}(\boldsymbol O)^2\right]\right)$, with $\boldsymbol O$ denoting a generic observation drawn from the true data-generating distribution. Accordingly, a consistent estimator of the variance is given by the empirical plug-in estimator
\[
\widehat{\mathrm{Var}}(\hat{\Delta})
=
\frac{1}{N^2}
\sum_{i=1}^N
\widehat{\mathrm{IF}}_{\Delta}(\boldsymbol O_i)^2,
\]
where $\widehat{\mathrm{IF}}_{\Delta}(\boldsymbol O_i)$ is obtained by evaluating $\mathrm{IF}_{\Delta}(\boldsymbol O_i)$ at consistent estimators of the unknown population quantities, including $\hat{\boldsymbol\theta}$.

We now connect the classical semiparametric theory above to modern neural network estimation by focusing on the regularized LBC-Net estimator. Because the propensity score is learned via a neural network with algorithm-dependent regularization (e.g., early stopping and ridge-type stabilization), the estimator converges to a regularized limit $\boldsymbol\theta^\star$ rather than a hypothetical unregularized true parameter $\boldsymbol\theta_0$. The resulting influence function, therefore, targets local variability around $\boldsymbol\theta^\star$, aligning inference with the implemented estimator.

For the influence function in \eqref{eq:if_delta_decomp} under $\boldsymbol\theta^\star$ to be valid, we impose three additional conditions that are compatible with neural-network architectures and training procedures. First, we assume the local identifiability and differentiability of the fitted network in a neighborhood of $\boldsymbol\theta^\star$, which holds for commonly used piecewise-smooth architectures (e.g., networks with ReLU activation functions; \citealp{tristan19, farrell2021deep}). Second, as a technical regularity condition, we assume local stability of the converged training solution, in the sense that small perturbations of the empirical distribution induce small changes in the stationary point reached by stochastic optimization algorithms (e.g., SGD or Adam); related stability properties of such methods are studied in \citet{kuzborskij2018data} and \citet{bocl2019}. Third, we assume that the empirical Jacobian of the training objective admits a well-conditioned (possibly regularized) inverse, ensuring the validity of a first-order Taylor expansion around the converged parameters, as is standard in influence-function analyses of trained deep networks \citep{koh2017}.

While influence functions in the deep-learning literature are typically interpreted as measures of prediction or parameter-level sensitivity to infinitesimal reweighting of training data \citep{cook1986localinfluence, koh2017, huang2023, lesci-etal-2024-causal}, they rely on the curvature of the empirical training loss and can be numerically unstable or difficult to interpret inferentially in high-dimensional settings \citep{basu2020influence, bae2022ifquestion, jimenez2025epistemic}. In contrast, LBC-Net targets a population-level statistical functional instead of the sensitivity of prediction, and its influence function admits an infinitesimal jackknife interpretation. Specifically, $\mathrm{IF}_{\boldsymbol\theta}(\boldsymbol X_i)$ characterizes how the propensity score weighted estimator responds to an infinitesimal perturbation of the $i$-th observation through the training procedure. The gradient $\partial \Delta_N(\boldsymbol\theta)/\partial \boldsymbol\theta^{\top}$ evaluated at $\boldsymbol\theta^\star$ then determines how these nuisance-parameter perturbations propagate into the target estimand $\Delta$. As a result, the contribution of nuisance estimation uncertainty enters the influence function only through a low-dimensional projection onto the functional-relevant direction. The remaining term, $\mathrm{IF}_{\mathrm{\Delta\mid \boldsymbol\theta^\star}}(\boldsymbol O_i)$, captures sampling variability while holding the nuisance parameter fixed. 

\textbf{Remark 3.1}. Analogous plug-in influence functions $\mathrm{IF}_{\mathrm{\Delta\mid \boldsymbol\theta_0}}(\boldsymbol O_i)$ arise for other IPTW-based estimands considered in this work. In particular, for time-to-event outcomes, we consider IPTW-weighted Nelson-Aalen estimators of the marginal survival functions under treatment and control, leading to survival and risk difference estimands of the form $S_1(t)-S_0(t)$ at fixed time points $t$ \citep{Mao2018surv}. The corresponding plug-in influence functions are derived under the IPTW-weighted Nelson-Aalen framework of \citet{DengWang2025} and are provided in the Supplementary Material (Section S1.1).

\textbf{Remark 3.2}. Under the correct specification of the propensity score model, the proposed variance estimator coincides with the classical sandwich variance for two-step estimators; see Supplementary Material (Section S1.2) for a formal derivation. This establishes compatibility with standard semiparametric variance theory in correctly specified parametric settings.\\

We now turn our attention to the numerical computation of the proposed variance estimator above, focusing on the high-dimensional parameters in the nuisance propensity score model. By plugging in the estimated propensity scores, the influence-function component $\mathrm{IF}_{\mathrm{\Delta\mid \hat{\boldsymbol\theta}}}(\boldsymbol O_i)$ can be computed directly. Moreover, backpropagation in neural networks makes the gradient $\partial \Delta_N(\boldsymbol\theta)/\partial \boldsymbol\theta^{\top}$ with respect to the fitted parameters computationally accessible via automatic differentiation, enabling the efficient evaluation of the chain-rule term. In practice, computing $\mathrm{IF}_{\boldsymbol{\theta}}(\boldsymbol{X}_i)$ is challenging because $G_0$ is a high-dimensional Jacobian of size $K(M+1)\times d_\theta$ ($d_\theta$ is the number of network parameters), making $G_0^{\top}G_0$ often ill-conditioned in neural network settings. The Jacobian $G_0$ is computed via automatic differentiation through the moment functions. Explicitly inverting $G_0^{\top}G_0$ is therefore computationally infeasible and numerically unstable, as naive inversion amplifies noise along weakly identified directions. To stabilize computation, we employ ridge-type regularization and approximate the inverse operator by $(G_0^{\top}G_0 + \tau I)^{-1}$, an approach similar to \citet{koh2017}. The inverse operator is approximated using a truncated singular value decomposition (SVD) of the empirical Jacobian. Writing $G_0 = U \Lambda V^\top$, we retain singular directions corresponding to sufficiently large singular values, defined via a relative spectral threshold $\Lambda_j \ge c \cdot \max_k \Lambda_k$ with $c = 10^{-3}$. Within this truncated subspace, we apply ridge regularization, replacing $\Lambda_j^{-2}$ with $(\Lambda_j^2 + \tau)^{-1}$. To adapt regularization to the local curvature of the estimating equations, we set $\tau = \alpha\,\mathrm{tr}(\Lambda^2)/d_\theta$ \citep{Grosse2023StudyingLL}. We use $\alpha = 0.01$ by default, which provides stable variance estimates. Sensitivity analyses over $\alpha \in [10^{-4}, 10^{-2}]$ show little effect on coverage, suggesting that moderate regularization stabilizes the estimator without inducing excessive variance inflation (Table S21).

\section{Simulation Study}
\label{sec:simulation}
\subsection{Simulation Settings}
We evaluated the proposed method under the widely-used benchmark simulation setting of \citet{kang2007demystifying}, referred to as ``KS simulation'' in this paper. The data generation process was described in the Supplementary Materials (Section S4). The sample sizes were N = 5,000, 1,000, and 300, and the results are aggregated from 500 Monte Carlo repetitions. Variance estimation and coverage diagnostics are evaluated separately under multiple sample sizes to illustrate moderate- and large-sample behavior. The estimand of interest in the KS simulation is the population mean of the outcome, $\mu=E(Y)$. Estimation of this mean under covariate-dependent missingness is mathematically analogous to ATE estimation in a two-group causal estimation framework \citep{kang2007demystifying}.

We compared the proposed method with the following approaches, which represent different ideas from the statistics and computer science literature: logistic regression, Covariate Balancing Propensity Scores (CBPS; \citealp{imai2014covariate}), standard neural network for binary outcomes (NN), neural network with imbalance score (NNIS; \citet{shang2025robust}), and Covariate Balancing with Integral Probability Metric (CBIPM; \citealp{kong2023covariate}). Logistic regression is the simplest and most widely used PS model in practice. CBPS, available in the R package \texttt{CBPS}, is one of the earliest and most widely used methods that use covariate balance constraints. The NN, which uses the binary cross-entropy (BCE) loss, represents off-the-shelf nonparametric machine learning methods for binary outcomes (e.g., random forest, kernel regression). It was implemented in Python using the \texttt{PyTorch} library. The NNIS represents another loss function that facilitates nonparametric PS estimation with balance constraints. The CBIPM represents a recent development in the machine learning literature on flexible weights that are not explicitly motivated by a PS model but leverage a neural network to minimize the loss that quantifies the overall difference in covariate distributions between treatment groups. The NNIS and CBIPM were implemented using the Python code disseminated with their publications, along with the suggested settings. We also used IPTW weights calculated from the true propensity scores as a benchmark, referred to as the ``true PS'' method. 

We evaluated the methods based on both covariate balance and outcome estimation. The former was evaluated using the GSD and LSD (PS-based methods only) for each covariate. The latter was evaluated using percent bias, root mean squared error (RMSE), and empirical standard deviation (ESD). 

In addition to the main simulation, we conducted extensive studies by expanding multiple simulation scenarios from \citet{li2021propensity}, focusing on poor overlap, heterogeneous treatment effects, and high-dimensional covariates. Scenarios with poor overlap are often characterized by many propensity scores near 0 or 1, leading to extreme weights. This is a widely recognized difficulty in the propensity score literature. We examined both homogeneous and heterogeneous effect scenarios. The latter allows individual causal effects to vary with propensity scores, which are more difficult to estimate due to their sensitivity to inappropriate propensity score estimation. Performance was evaluated across multiple sample sizes, with representative results reported for N=5,000, which mirror our data application with 20 covariates. To assess LBC-Net's performance in high-dimensional settings, we also simulated data with 30,000 samples and 84 covariates. We also conducted experiments to evaluate robustness to tuning parameters. These results are presented in the Supplementary Materials due to space constraints, but their main conclusions are summarized at the end of this section.

\subsection{Simulation results on covariate balance and treatment effect estimation}

\begin{figure}[ht]
\centering
\includegraphics[width=0.9\textwidth]{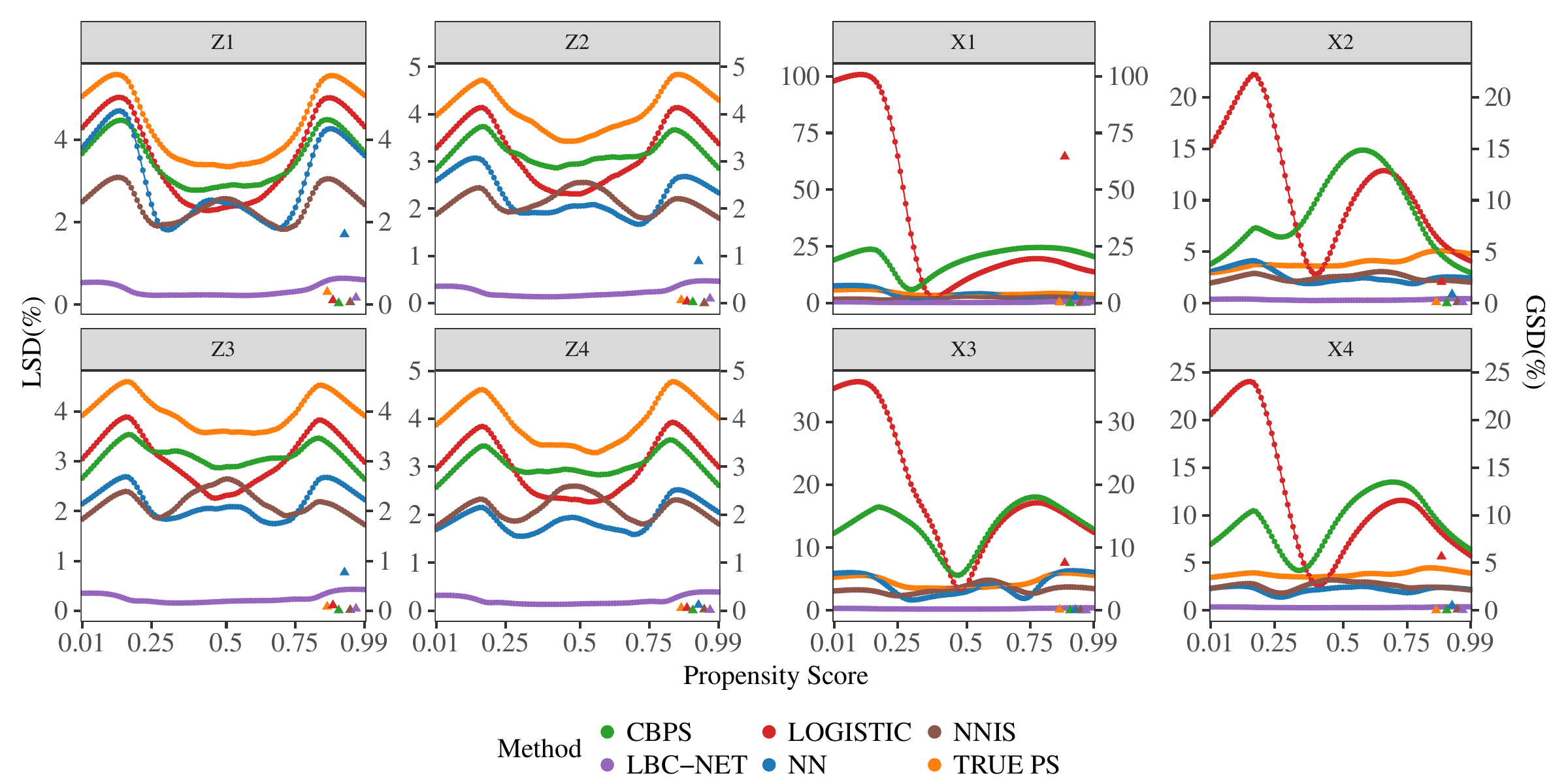}
\caption{
Local and global standardized differences (LSD and GSD) of covariates for four model-based propensity score methods under the KS simulation setting. The left panel corresponds to the standard data-generating process with covariates $Z_1$-$Z_4$, and the right panel to the extended process with transformed covariates $X_1$--$X_4$. LSD curves are plotted on the left vertical axis over a grid from 0.01 to 0.99, and GSD values are shown as triangles on the right axis.
}
\label{fig:KS5k.lsd}
\end{figure}

We first present results with $N=5,000$ from the KS simulation, which includes two data-generating processes (DGPs): the Standard DGP, where the PS model is a correctly specified logistic regression with main effects; and the Extended DGP, which involves complex nonlinear transformations unknown to the analyst, leading to model misspecification. Under the Standard DGP, all methods achieved low GSD regardless of whether they relied on parametric assumptions (e.g., Logistic, CBPS; Figure \ref{fig:KS5k.lsd}). As expected, logistic regression performs well here due to correct model specification. For local balance, LBC-Net consistently outperformed other model-based approaches, reinforcing its suitability for PS estimation. Note that the LSD is generally higher than the GSD due to smaller local sample sizes. Crucially, Table \ref{tab:KS.Table5k} shows that LBC-Net achieved population mean estimates that are at least as accurate and efficient as those obtained by parametric methods, even under correctly specified models—a significant result given the difficulty of model diagnosis in IPTW, as discussed in Section \ref{sec:intro}. This result supports using LBC-Net as an automated PS  procedure in all situations, regardless of whether parametric assumptions hold. Other methods, such as NN, NNIS, and CBIPM, also performed well. Although NN and LBC-Net share similar neural network architectures, LBC-Net exhibited superior numerical stability, covariate balance, and efficiency (Figure \ref{fig:KS5k.lsd}, Table \ref{tab:KS.Table5k}), supporting our claim in Section \ref{subsec:nonpar.PS} that the proposed tailored loss is preferable to standard binary classification losses. 

\begin{table}[ht]
\centering
\caption{Results under the KS simulation setting with $n=5{,}000$.} % The true outcome population mean is $210$.
\begin{small}
\adjustbox{max width=\textwidth}{
\begin{tabular}{rrrrrrrrr}
\toprule
 &  & \multicolumn{3}{c}{Outcome: $\mu=\mathbb{E}\{Y\}$} & \multicolumn{4}{c}{Covariates: GSD} \\
 \cmidrule(lr){3-5}\cmidrule(lr){6-9} 
DGP & Methods & \% Bias & RMSE & ESD & X1 & X2 & X3 & X4 \\
\midrule 
 & True PS & -0.01 & 1.01 & 1.01 & -0.31 & 0.07 & -0.08 & -0.06 \\
 & Logistic & -4e-04 & 0.76 & 0.76 & -0.11 & 0.04 & -0.11 & -0.06 \\
 & NNIS & 0.05 & 0.58 & 0.57 & -0.06 &  1e-03 & 0.02 & -0.03 \\
Standard & CBIPM & 0.01 & 0.49 & 0.49 & -0.03 & -0.02 & -0.01 & -0.02 \\
 & CBPS & 0.01 & 0.64 & 0.64 & -0.04 & 0.02 & -0.01 & -0.01 \\
 & NN & -0.15 & 0.66 & 0.59 & -1.71 & 0.89 & -0.77 & -0.12 \\
 & LBC-Net & -0.01 & 0.50 & 0.50 & -0.17 & 0.10 & -0.04 & -0.02 \\
\midrule 
 & True PS & -0.01 & 1.01 & 1.01 & -0.34 & 0.13 & -0.11 & 0.01 \\
 & Logistic & 4.22 & 16.75 & 14.23 & 64.59 & 2.07 & -7.57 & 5.65 \\
 & NNIS & -0.25 & 0.76 & 0.55 & -0.10 & -0.11 & -0.01 & 0.07 \\
Extended & CBIPM & -0.35 & 0.95 & 0.61 & 0.01 & -0.04 & -8e-04 & 0.01 \\
 & CBPS & -0.78 & 1.80 & 0.77 & -0.03 & 0.02 & 0.02 & -1e-03 \\
 & NN & -0.06 & 2.58 & 2.58 & 2.90 & 0.87 & -0.08 & 0.51 \\
 & LBC-Net & -0.46 & 1.11 & 0.56 & -0.20 & 0.11 & -0.03 & 0.03 \\
 \bottomrule
\end{tabular}
}
\end{small}
\label{tab:KS.Table5k}
\end{table}

When the parametric model is misspecified (Extended DGP; Figure \ref{fig:KS5k.lsd}), both Logistic and CBPS exhibit large LSD, as expected due to their reliance on a misspecified PS model. Although CBPS enforces near-zero GSD by design, the LSD raises concerns about whether such global balance masks underlying model misspecification, and whether the resulting propensity scores and their inverse probability weights retain their desired properties—an issue that becomes more pronounced in the more complex designs (Table S18 and Figure S25). In contrast, the nonparametric methods, which are less sensitive to model misspecification, generally perform better, with LBC-Net achieving improved efficiency and numerical stability relative to standard neural network–based estimation (NN) while maintaining bias comparable to other balancing approaches. This superior performance can be attributed to its enhanced numerical stability, owing to the use of a large number of balance constraints across $K$ grid points. Notably, numerical stability in covariate balance (reflected in GSD and LSD) and in outcome estimation (reflected in bias and efficiency) is closely linked, as both are influenced by IPTW weights in a similar manner. LBC-Net demonstrates consistent robustness across all simulation settings and the data application. In particular, Figure \ref{fig:KS5k.lsd} shows that while other methods exhibit high LSD near 0 or 1 (where data are sparse), LBC-Net consistently maintains low LSD across all grid points. Furthermore, the local calibration plot (Figure S11) confirms that LBC-Net satisfies local calibration constraints. Results for smaller sample sizes ($N = 1,000$ and $300$) are reported in Supplementary Material Section S4. The relative performance of the methods is broadly similar to or better than that of others.

The Supplementary Materials present additional simulation studies supporting the robustness of LBC-Net. First, LBC-Net consistently performed well across all settings, avoiding model misspecification, optimizing covariate balance, and yielding stable and efficient IPTW estimates with negligible bias. 

Second, it remained effective in more challenging scenarios with poor overlap and heterogeneous treatment effects, particularly with larger sample sizes (Supplementary Sections S5.2 and S6). Some balancing-based methods, such as NNIS, may suffer from numerical instability in these regimes (Supplementary Section S6). CPIPM, a representative of recently developed methods that directly balance multivariate covariate distributions, exhibited deteriorated performance in such challenging settings (Supplementary Section S5.2).

Third, with a fixed number of covariates in $\boldsymbol Z$, increasing the number of neighborhoods $K$ in the loss function improves estimation efficiency until it reaches a plateau (Supplementary Materials Section S9). Since $K$ determines the number of local balance constraints, this finding suggests that our default $K = 99$ is a conservative and effective choice, while using only global balance ($K = 1$) is insufficient—offering a perspective on LBC-Net's superior stability and efficiency compared to some global balance-based methods. 

Fourth, in the KS simulation, we observed that local balancing of first moments also improves balance in second moments (Supplementary Materials Section S8). Although such improvements are not guaranteed in general, the LBC-Net framework permits $\boldsymbol Z$ to incorporate user-specified transformations of the covariates (e.g., nonlinear terms or interactions), thereby enabling balance constraints to operate on richer representations of the covariate distribution. This flexibility is supported in our \texttt{R} package through user-defined specification of $\boldsymbol Z$, and it is illustrated empirically in the data application of Section~\ref{sec:application} and Supplementary Section S2.

\subsection{Simulation results on variance estimation}

The full simulation results on the comparison between the proposed IF-based variance estimator and the bootstrap are included in the Supplementary Materials Section S7. Table~\ref{tab:KSse.Table5k} presents selected results under the KS standard DGP to demonstrate the representative performance. The IF-based standard errors closely match both the empirical sampling variability of the IPTW estimator and the bootstrap estimates. In particular, the SD/SE ratio remains close to one across all scenarios, ranging from 0.82 to 1.17, indicating accurate variance calibration. Overall, the two methods yield highly comparable results. 

Regarding confidence interval coverage, near-nominal 95\% coverage was achieved in all scenarios except the misspecified KS design, where coverage fell below nominal levels for all methods (Supplementary Section S7). As discussed earlier, this behavior reflects a small but systematic first-order bias in the IPTW estimator rather than variance miscalibration. Consistent with this interpretation, SD/SE ratios remained close to one even in this setting, indicating that both the proposed variance estimator and the bootstrap accurately captured estimator variability. The resulting under-coverage arises because estimator variability is low in this design, so even modest bias can materially affect coverage.

Taken together, these results indicate that the proposed IF-based variance estimator matches the bootstrap in statistical accuracy while offering substantial computational advantages. A refit bootstrap requires repeated model fitting and therefore incurs a computational cost of order $\mathcal{O}(B \cdot C_N)$, where $B$ denotes the number of bootstrap resamples and $C_N$ is the cost of fitting the propensity score model once. In contrast, the proposed estimator requires only a single model fit, with variance computed via a limited number of additional backpropagation operations. Although this introduces some overhead relative to point estimation, the cost does not scale with B, and remains substantially lower than that of a refit bootstrap. Additionally, the theoretical validity of bootstrapping in deep neural network methods remains insufficiently established.

\begin{table}[ht]
\renewcommand{\arraystretch}{1.15}
\small
\centering
\caption{Comparison of variance estimation methods under the KS standard DGP. Reported are the Monte Carlo mean (over 500 repetitions) of the point estimator (Mean Est), empirical standard deviation (SD Est), average estimated standard error (Mean SE), empirical 95\% coverage (Coverage), and the ratio SD/SE.}
\vskip 0.15in
\begin{tabular*}{\textwidth}{@{\extracolsep\fill}lrrcccccc}
\toprule
& & & \multicolumn{3}{c}{IF-based} & \multicolumn{3}{c}{Bootstrap} \\
\cmidrule(lr){4-6} \cmidrule(lr){7-9}
N & True Effect & Mean Est
& Mean SE & Coverage & \textbf SD/SE
& Mean SE & \textbf Coverage & SD/SE \\
\midrule

\multirow{4}{*}{}
500   & 210 & 209.99 & 1.69 & 0.97 & 0.88 & 1.61 & 0.96 & 0.93 \\
2000  & 210 & 209.98 & 0.87 & 0.96 & 0.91 & 0.82 & 0.95 & 0.97 \\
5000  & 210 & 210.00 & 0.55 & 0.96 & 0.91 & 0.51 & 0.96 & 0.98 \\
10000 & 210 & 209.94 & 0.40 & 0.95 & 0.92 & 0.37 & 0.95 & 1.00 \\

\bottomrule
\end{tabular*}
\label{tab:KSse.Table5k}
\end{table}

\section{Illustrative Data Application}
\label{sec:application}

We illustrate the proposed methodology with a dataset from the MIMIC-IV electronic healthcare database version 3.0 \citep{johnson2024mimic}, which includes sepsis patients admitted to intensive care units (ICUs) between 2008 and 2022. For sepsis patients, the erythrocyte distribution width to platelet ratio (EPR) is a key biomarker with significant prognostic value \citep{yao2023red}. While EPR is a physiological marker and is not a directly modifiable treatment, its trajectory over time serves as a proxy for the effectiveness of the underlying clinical management in controlling inflammation. Therefore, we aim to estimate the causal effect of being on a physiological path corresponding to unsuccessful versus successful inflammatory response management, justifying the study design and scientific rigor of our case study.

The study included $5,518$ ICU patients with a minimum stay of 96 hours, excluding those discharged or deceased earlier. Day 4 (96 hours post-ICU admission) marked the beginning of outcome observation and separated baseline confounders from post-treatment outcomes. The exposure was the relative percent change from the baseline EPR (the first measurement $\geq$24 hours post-admission) to Day 3 EPR (the first value within the 72–96 hour window). Patients were categorized as having elevated ($T=1$) or stable ($T=0$) EPR using a $30\%$ threshold per \citet{zhou2022enhanced}. The primary outcome was marginal survival through 7, 14, and 28 days, measured from the end of Day 3; the corresponding estimands are survival probability differences at these fixed horizons. Secondary outcomes included in-hospital mortality and hospital length of stay, with the latter truncated at 180 days for seven patients with prolonged admissions. Baseline confounders comprised demographic characteristics, admission type, vital signs, laboratory measurements, and early ICU interventions, yielding a total of 19 covariates selected based on clinical relevance and data availability. Additional details on eligibility criteria, variable definitions, and study design are provided in the Supplementary Materials (Section S2).

\begin{figure}[ht]
\centering
\includegraphics[width=0.9\columnwidth]{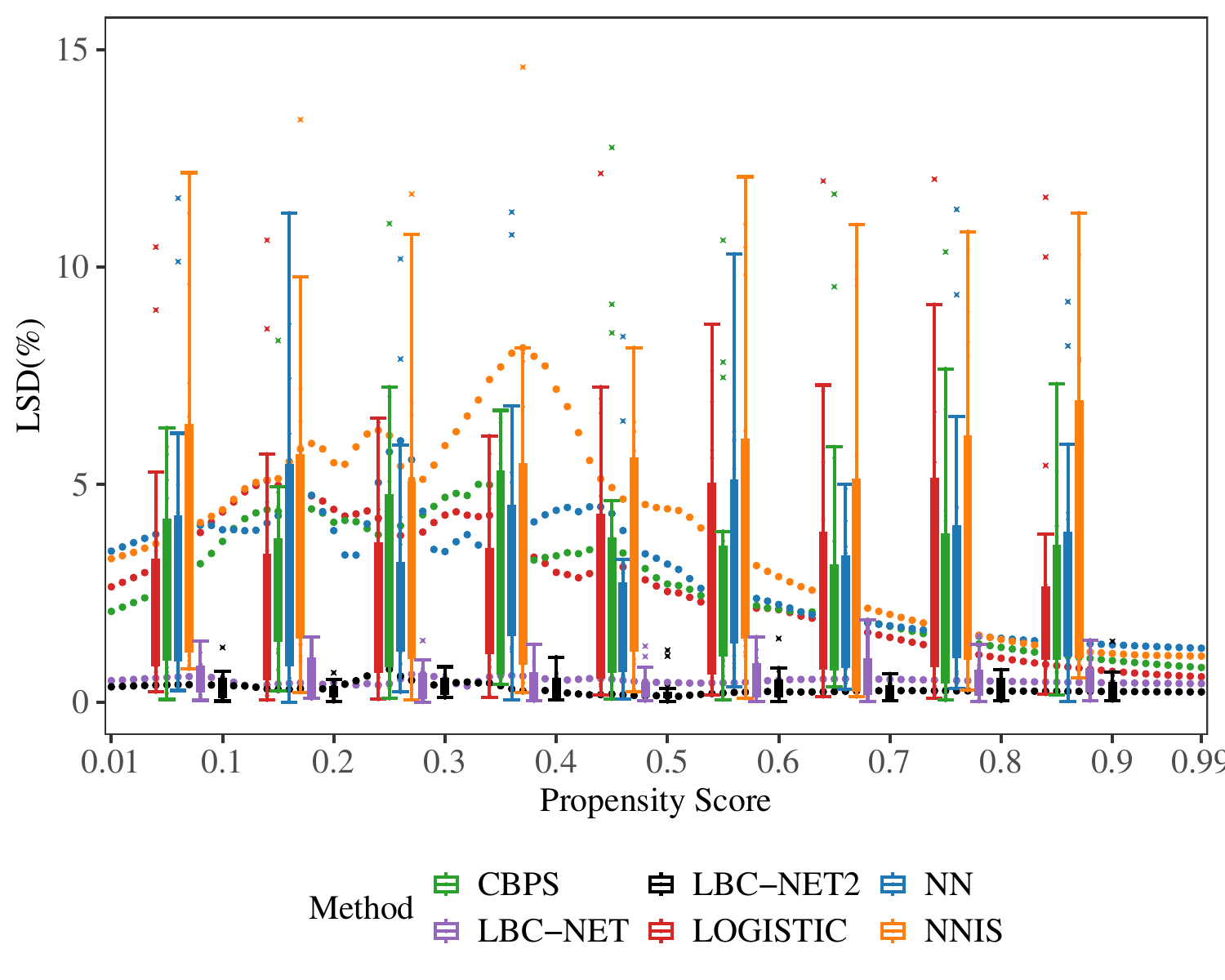}
\caption{The LSD of 19 covariates in the analysis of MIMIC-EPR data. Each dot on the curves represents the average LSD over all the covariates at the corresponding grid point of the propensity score, drawn to the vertical axis on the left. The four methods are symbolized by different colors. The boxplots show further details of the LSD of individual covariates at selected local neighborhoods, revealing their ranges, dispersion and outliers (denoted by cross symbols). The LBC-Net has uniformly better local covariate balance at all levels of the propensity scores and for all covariates, which is also a manifestation of its stability of estimation. The CBIPM method does not use a propensity score and hence cannot be presented in this plot.}
\label{fig:mimic.lsd}
\end{figure}

We applied all methods used in the simulation study to this real dataset. Following standard practice \citep{austin2015moving}, we report GSDs before and after PS adjustment (Tables S1 and S2). In addition to the baseline LBC-Net specification, which balances the original covariates, we also consider an extended version that augments the balance constraints with quadratic terms and pairwise interactions of continuous covariates; this variant is referred to as LBC-Net2. All methods improved global balance, with LBC-Net and LBC-Net2 exhibiting the most favorable balance overall. Even methods without explicit balance constraints (e.g., Logistic and NN) achieved substantial imbalance reduction (GSD $<5\%$). Notably, this dataset exhibited good overlap between treatment groups (Figure S4), providing favorable conditions for propensity score weighting adjustment. However, the local balance analysis uncovered hidden problems. LBC-Net consistently achieved low LSD (Figure \ref{fig:mimic.lsd}), with minimal between-covariate variation and no outliers in boxplots, indicating uniform balance across all 19 covariates. In contrast, other methods showed much larger LSD and more variation and outliers. Local calibration analysis revealed notable deviations from expected PS performance for parametric models (Figure S3), questioning the validity of their IPTW. Traditional diagnostics further showed nonlinear relationships between continuous covariates and treatment (Figure S6). While adding interactions or nonlinear terms may help, model tuning is challenging due to high dimensionality and a lack of clear guidance; the possible instability in IPTW weights may also mislead balance-based model refinement, but the effect is difficult to evaluate. Despite these limitations, logistic regression with main effects remains the predominant approach in medical publications \citep{Zhang2019diagnosis}. 

We therefore evaluated whether explicitly incorporating higher-order balance constraints improves performance. LBC-Net2 achieved a systematically tighter balance than LBC-Net, as reflected in improved LSDs (Figure \ref{fig:mimic.lsd}), reduced GSDs (Table S2), and enhanced calibration diagnostics (Figure S3). These improvements are consistent with the stronger moment conditions imposed by the expanded feature space. To further assess distributional balance, we compared empirical covariate distributions using the Kolmogorov–Smirnov $D$ statistic; results reported in the Supplementary Materials indicate that LBC-Net and LBC-Net2 attain a more favorable distributional balance than competing methods (Table S3).

For the survival outcome, our primary analysis focuses on marginal survival probability differences at fixed time horizons, which permit IF-based variance estimation. Hazard ratios from Cox models are reported as a secondary analysis for clinical comparability and are deferred to the Supplementary Materials. Using LBC-Net, elevated EPR was associated with a statistically significant reduction in survival probability at all horizons, with the largest effect observed at 14 days ($-8.1$ percentage points; 95\% CI: [$-11.8$, $-5.3$]) and a still substantial effect at 28 days ($-6.3$ percentage points; 95\% CI: [$-10.0$, $-2.6$]). Table~\ref{tab:mimic.Table} presents estimates across alternative methods that are qualitatively similar and reflect the favorable overlap structure of the cohort. LBC-Net exhibited the most stable inference alongside superior balance and calibration diagnostics. For LBC-Net and LBC-Net2, uncertainty was quantified using the proposed IF-based variance estimator, whereas bootstrap variance estimates were used for competing methods; bootstrap results for LBC-Net were highly consistent with IF-based estimates (Table~S5).

For the secondary outcomes of in-hospital mortality and hospital length of stay, effect estimates were small and not statistically significant across all methods, indicating limited evidence of a causal impact of short-term EPR dynamics on these outcomes in this cohort.

\textbf{Remark 4.1}. When multiple PS methods are applied to a dataset, they often yield numerically different ATE estimates, sometimes with substantial discrepancies. Which estimates should we trust? In real-world analyses, bias and efficiency are unknown, and findings from prior studies may not generalize. We advocate trusting methods that yield ``correct'' propensity scores, evaluated without referencing the outcomes \citep{Rubin2008}. Proposition \ref{thm:two.conditions} provides the diagnostic criteria for this ``correctness'': local balance and local calibration, which can be visualized via LSD plots (e.g., Figures \ref{fig:KS5k.lsd} and \ref{fig:mimic.lsd}) and calibration plots (e.g., Figure S3), respectively. This approach is more rigorous than relying on GSD alone, as near-zero GSD may mask local imbalances and hide model misspecification. Furthermore, because these two conditions are necessary and sufficient, they also provide a more complete assessment than regression model diagnostics, such as those used for logistic regression.

\begin{table}[ht]
\centering
\caption{\label{tab:mimic.Table}
Estimation of the average treatment effect (ATE) for hospital length of stay (LOS),ATE for in-hospital mortality, and survival differences at 7, 14, and 28 days in the MIMIC-IV dataset. Results are reported as point estimates with 95\% confidence intervals (CI). Survival differences and mortality effects are expressed in \textbf{percentage points (pp)}. }
\fontsize{10}{12}\selectfont
\resizebox{\ifdim\width>\linewidth\linewidth\else\width\fi}{!}{%
\begin{tabular}{l p{2.8cm} p{3.0cm} p{2.8cm} p{2.8cm} p{2.8cm}}
\toprule
 & & & \multicolumn{3}{c}{Survival Difference (\textbf{pp})} \\
\cmidrule(lr){4-6}
Method & LOS & Mortality (\textbf{pp}) & 7 days & 14 days & 28 days \\
\midrule

Logistic &-0.17 (0.52)\newline
[-1.20, 0.85] & 10.1 (1.5)\newline
[7.2, 13.0] & -5.1 (0.9)\newline
[-6.8, -3.4] & -8.0 (1.4)\newline
[-10.8, -5.3] & -6.6 (1.9)\newline
[-10.2, -2.9]\\

CBIPM & -0.58 (0.51)\newline
[-1.58, 0.42] & 10.1 (1.5)\newline
[7.1, 13.0] & -5.1 (0.9)\newline
[-6.9, -3.3] & -8.5 (1.5)\newline
[-11.4, -5.6] & -7.6 (2.0)\newline
[-11.4, -3.7]\\

CBPS & -0.20 (0.52)\newline
[-1.22, 0.81] & 10.2 (1.5)\newline
[7.3, 13.1] & -5.1 (0.9)\newline
[-6.9, -3.4] & -8.2 (1.4)\newline
[-10.9, -5.4] & -6.7 (1.8)\newline
[-10.3, -3.1]\\

NN & -0.21 (0.95)\newline
[-2.07, 1.64] & 9.7 (1.9)\newline
[5.9, 13.5] & -4.8 (1.3)\newline
[-7.3, -2.3] & -7.9 (1.8)\newline
[-11.4, -4.5] & -6.1 (2.4)\newline
[-10.8, -1.3]\\

NNIS & -0.11 (0.56)\newline
[-1.21, 1.00] & 10.1 (1.5)\newline
[7.1, 13.1] & -4.9 (1.0)\newline
[-6.8, -3.1] & -7.8 (1.5)\newline
[-10.8, -4.9] & -6.4 (2.0)\newline
[-10.4, -2.5]\\

LBC-net & -0.13 (0.51)\newline
[-1.14, 0.88] & 10.0 (1.5)\newline
[7.2, 12.9] & -5.1 (1.0)\newline
[-7.0, -3.2] & -8.1 (1.4)\newline
[-11.0, -5.3] & -6.3 (1.9)\newline
[-10.0, -2.6]\\

LBC-net2 & -0.06 (0.54)\newline
[-1.13, 1.01] & 10.4 (1.5)\newline
[7.5, 13.3] & -4.5 (0.9)\newline
[-6.3, -2.6] & -8.6 (1.5)\newline
[-11.5, -5.7] & -7.3 (1.9)\newline
[-11.0, -3.6]\\
\bottomrule
\end{tabular}
}
\end{table}

\section{Discussion and conclusion}
\label{sec:discussion}

Machine learning methods can be used for both propensity score and outcome modeling, enabling doubly robust estimation \citep{2011TMLEbook, Chernozhukov2018, 2019dragonnet}. As a deep learning model with a tailored loss function, LBC-Net can also be extended to incorporate outcome models. However, this paper focuses on covariate balance, and a comparison with augmented methods is beyond its scope. In the typical practical situation where a single real dataset is analyzed, the true bias and efficiency of an estimator are unknown, and different methods often yield varying results. Among them, those achieving better covariate balance are generally preferred (Remark 4.1). Therefore, regardless of whether outcome models are used, covariate balance remains essential for the validity of PS-based analysis. Since outcome models heavily influence bias and efficiency, we exclude augmented estimators to isolate the impact of covariate balance on treatment effect estimation. A thorough discussion of augmented estimation across diverse outcome types is valuable and will be pursued in future work.

Accordingly, the proposed variance estimator differs from efficient influence function–based inference in doubly robust frameworks \citep{2011TMLEbook, Chernozhukov2018}. Those approaches use orthogonal estimating equations involving both propensity score and outcome models to remove first-order sensitivity to nuisance estimation error. In contrast, our setting is based on the IPTW estimator, and no orthogonalization is imposed; the uncertainty from propensity score estimation is propagated directly through the influence function.

Our empirical analysis highlights a key theoretical insight: although the true propensity score minimizes covariate imbalance asymptotically, it yields a non-zero balance loss in finite samples due to sampling variability. In contrast, the estimated propensity scores, optimized with covariate balance constraints, achieve near-zero balance loss and smaller finite-sample bias in estimating the ATE. This aligns with \citet{Rosenbaum1987} influential finding that estimated propensity scores can outperform true scores in bias reduction and balance—an observation widely validated in the literature. The counterintuitive advantage arises because balance-driven estimation explicitly corrects for finite-sample imbalance, enhancing practical performance even at the cost of deviation from the true score.

The local balance equation $E[ T W \boldsymbol Z | S ] - E[ (1-T) W \boldsymbol Z | S ] = \boldsymbol 0 $ provides a necessary but not sufficient condition for achieving conditional independence between $T$ and $\boldsymbol Z$ given $S$. While we have observed in the numerical studies that balancing the first moments of covariates also helps improve the balance of the second moments (see Supplementary Materials Section S8) and empirical cumulative distribution functions (Section \ref{sec:application}), future research should focus on developing more comprehensive local balance constraints for the entire covariate distribution, similar to the recent developments in the global balance constraint methods \citep{Covariate.distribution.balance.2022, kong2023covariate}. Of note, $\boldsymbol D_{1k}$ suggests that the local balance constraint can be interpreted as a global balance constraint on $\omega(c_k, p_j) \boldsymbol Z_j$, a special function of the covariates, for $k=1,2,..., K$. It is of interest to study the relationship between the large number of constraints used in our paper and the large number of constraints employed in some global balance methods aimed at balancing the distribution. 

This paper focuses on PS model specification and covariate balance in order to achieve accurate ATE estimation, with the concepts of local balance and local calibration playing key roles. However, these concepts are not always indispensable. First, in a doubly robust procedure, covariate balance is dispensable if the outcome model is correctly specified. Second, when the outcome is linearly related to covariate features, mean balance in those features suffices for consistent ATE estimation, even without a propensity score model \citep{hazlett2018kernel}. Third, weights that balance the joint covariate distribution need not be derived from propensity scores \citep{benmichael2021balancing}. Still, reporting covariate balance before and after propensity score adjustment remains standard practice \citep{austin2015moving}, and balance diagnostics are valuable for selecting reliable methods (Remark 4.1). Since the outcome and propensity score models regress different dependent variables on the same set of covariates, if a nonparametric algorithm is used for estimating these models, they may have some misspecification simultaneously. Therefore, covariate balance and stable weight calculation are especially important in data analytical practice \citep{kang2007demystifying}. Satisfying the sufficient and complete conditions of Proposition 2.1 is also very important for validity. Our proposed method demonstrates superior numerical performance and properties in these regards.

\section*{Acknowledgments}
This research is partially supported by NIH grants R01CA225646, R01HL175410, and P30CA016672, and CPRIT grant RP210130. 

\section*{Conflict of interest statement}
The author(s) declared no potential conflicts of interest with respect to the research, authorship, and/or publication of this article.

\section*{Data Availability Statement}
The MIMIC-IV data is publicly available at: https://physionet.org/content/mimiciv/3.0/. Data access requires credentialed approval, a data use agreement, and training. 

%Bibliography
\bibliographystyle{apalike}  
\bibliography{references}

\end{document}